\documentclass[11pt]{article}
\usepackage{moriond,epsfig}
\usepackage{array}
\def\Journal#1#2#3#4{{#1} {\bf #2}, #3 (#4)}

\def\NIMA{{\em Nucl. Instrum. Methods} A}

\def\PRL{\em Phys. Rev. Lett.}
\def\PRD{{\em Phys. Rev.} D}

\def\be{\begin{equation}}
\def\ee{\end{equation}}
\def\bea{\begin{eqnarray}}
\def\eea{\end{eqnarray}}

\begin{document}
\vspace*{4cm}
\title{BEAUTY AND CHARM PRODUCTION CROSS SECTION MEASUREMENTS AT THE TEVATRON}

\author{ J.M.~PURSLEY \\ on behalf of the CDF and D\O~Collaborations }

\address{Physics Department, Room 2320 Chamberlin Hall, University of Wisconsin-Madison,\\
1150 University Avenue, Madison, WI 53706, USA}

\maketitle\abstracts{
Heavy quark production probes QCD at the interface 
of the perturbative and non-perturbative regimes.
Studying the production of heavy quarks is an
important test of models in both regimes. In this
article, recent results on beauty and charm production 
from the CDF and D\O~experiments at the Tevatron are
reported.  These include measurements of correlated 
$b\bar{b}$ production, the $\psi(2S)$ production cross 
section, and  $\Upsilon(1S)$ and $\Upsilon(2S)$ 
polarization.
}

\section{Introduction\label{sec:intro}}

The study of heavy quark production is an important
test of models of both perturbative and non-perturbative
quantum chromodynamics
(QCD).  Since the development of nonrelativistic QCD
(NRQCD), the agreement between theoretical predictions
and measured values of heavy quark production cross
sections has improved greatly.  However, recent $J/\psi$
and $\psi(2S)$ polarization measurements at the Tevatron~\cite{psiPol}
indicate the understanding of heavy quark production is 
not yet complete.

The Tevatron at the Fermi National Accelerator Laboratory
produces $p\bar{p}$ collisions with a center of mass energy
of $1.96$ TeV.  The Collider Detector at Fermilab (CDF) and
D\O~experiments employ general multipurpose detectors~\cite{CDF,D0} 
to reconstruct particle physics events from these collisions.  
With $b$-hadron cross sections of $\sim$30 $\mu$b 
$\big(|\eta| < 1.0\big)$,~\cite{CDF,det_coord} CDF and 
D\O~have a wealth of experimental data on $b$-hadrons.

\section{Correlated $b\bar{b}$ Production\label{sec:bbbar}}

The cross section for producing, in hadronic collisions,
both the $b$ and $\bar{b}$ quarks centrally and above a 
given $p_{\rm T}$ threshold is referred to as the $b\bar{b}$ 
correlation $\sigma_{b\bar{b}}$.  The exact next-to-leading
order (NLO) prediction of $\sigma_{b\bar{b}}$ appears to
be a robust perturbative QCD prediction, differing by only
a few percent from the leading order (LO) prediction.
However, Run I measurements of $\sigma_{b\bar{b}}$ at
the Tevatron are inconclusive, with an average ratio of
the measured $\sigma_{b\bar{b}}$ over the exact NLO
prediction ($R_{2b}$) of $1.8$ with a $0.8$ RMS 
deviation.~\cite{run1bbbar}

In this measurement,~\cite{bbbar} $\sigma_{b\bar{b}}$ is 
obtained in $740$ pb$^{-1}$ of CDF data using dimuon 
events with $p_{\rm T}(\mu)>3$ GeV/$c$, $|\eta|<0.7$, 
and invariant mass $m_{\mu\mu} \in [5, 80]$ GeV/$c^2$.
This corresponds to $b\bar{b}$ pairs with $p_{\rm T} \ge 2$
GeV/$c$ and rapidity $|y| \le 1.3$.
At the Tevatron, dimuon events mainly result 
from the decay of heavy quark pairs ($b\bar{b}$ or 
$c\bar{c}$), the Drell-Yan process, the decay of 
charmonium and bottomonium, and the decay or
misidentif\mbox{}ication of $\pi$ or $K$ mesons.  
To determine the $b\bar{b}$ and $c\bar{c}$ content of 
the data, we f\mbox{}it the impact parameter 
distribution of the muon tracks.  The impact parameter 
$d_0$ is  def\mbox{}ined as the distance of closest 
approach of the track to the primary event vertex in 
the transverse plane, and it is proportional to the 
decay time of the parent particle.
 
The one-dimensional impact parameter distributions of 
muons from $b$ and $c$ decays are modeled by a tuned 
Herwig simulation,~\cite{bbbar} while the distributions of prompt 
muons are reconstructed using muons from $\Upsilon(1S)$ 
decays in the data.  Because the impact parameters of the
two muons are to first order uncorrelated, the three 1D 
templates may be combined into six 2D templates to represent 
each possible dimuon source ($b\bar{b}$, $c\bar{c}$, $cb$, 
prompt-prompt, prompt-$b$, and prompt-$c$).  These six
templates are then used to perform a maximum likelihood
f\mbox{}it to the 2D distribution of the impact parameter 
of both muons to extract the $b\bar{b}$ and $c\bar{c}$ components.  
The projection of the 2D impact parameter distribution 
is compared to the f\mbox{}it result in Figure \ref{fig:bbbar}.

From this f\mbox{}it, we measure the dimuon cross sections to be
$\sigma_{b\to\mu,\bar{b}\to\mu} = 1549 \pm 133$ pb and
$\sigma_{c\to\mu,\bar{c}\to\mu} = 624 \pm 104$ pb, where
the quoted error is the sum in quadrature of statistical
and systematic uncertainties.  In order to compare with
theoretical predictions, we evaluate the NLO dimuon cross
section using the {\sc MNR} generator with the events
decayed by EvtGen.~\cite{bbbar}
This gives a value 
of $\sigma_{b\to\mu,\bar{b}\to\mu}^{\rm NLO} = 1293$ pb,
resulting in a ratio of $R_{2b} = 1.2\pm0.2$.  This
measurement is in agreement with the NLO theoretical 
prediction, and does not conf\mbox{}irm the anomalously high
dimuon cross section observed in Run I.

\section{$\psi(2S)$ Production Cross Section\label{sec:psi2s}}

Charmonium production provides another arena in which to 
test our understanding of QCD.  The development of
NRQCD was prompted in part by the CDF Run I measurements 
of $J/\psi$ and $\psi(2S)$ production cross sections.~\cite{run1psi2s} 
A new CDF measurement of the $\psi(2S)$ production cross 
section pushes the $p_{\rm T}$ range farther into the
perturbative QCD regime than was possible with Run I
data.~\cite{psi2s}

We reconstruct $\psi(2S) \to \mu^+\mu^-$ using 1.1 fb$^{-1}$
of CDF data.  We then perform an unbinned maximum likelihood 
f\mbox{}it in the $\psi(2S)$ mass and proper decay length $ct$ 
distributions.  The mass f\mbox{}it separates the signal from the
background, while the $ct$ f\mbox{}it separates promptly-produced 
$\psi(2S)$ from $\psi(2S)$ originating from secondary 
decays of long-lived particles (predominantly $B$ mesons).
The $\psi(2S)$ $p_{\rm T}$ range of 2 to 30 GeV/$c$ is
divided into 25 bins, and the signal yield and prompt
fraction in each $p_{\rm T}$ bin are extracted by the 
likelihood f\mbox{}it.

The $\psi(2S)$ acceptance depends upon the $\psi(2S)$
polarization.  The CDF measurement of $\psi(2S)$
polarization is statistically limited, with only three 
measured data points.~\cite{psiPol}  We take the average
of these three data points as the effective polarization
$\alpha_{\rm eff} = 0.01 \pm 0.13$, where $\alpha$ is
def\mbox{}ined according to Equation (\ref{eq:alpha}),
and use this value to calculate the $\psi(2S)$ acceptance 
in each bin of $\psi(2S)$ $p_{\rm T}$.

\begin{figure}[tbh]
\begin{minipage}[tbh]{0.5\linewidth}
\centering
\psfig{figure=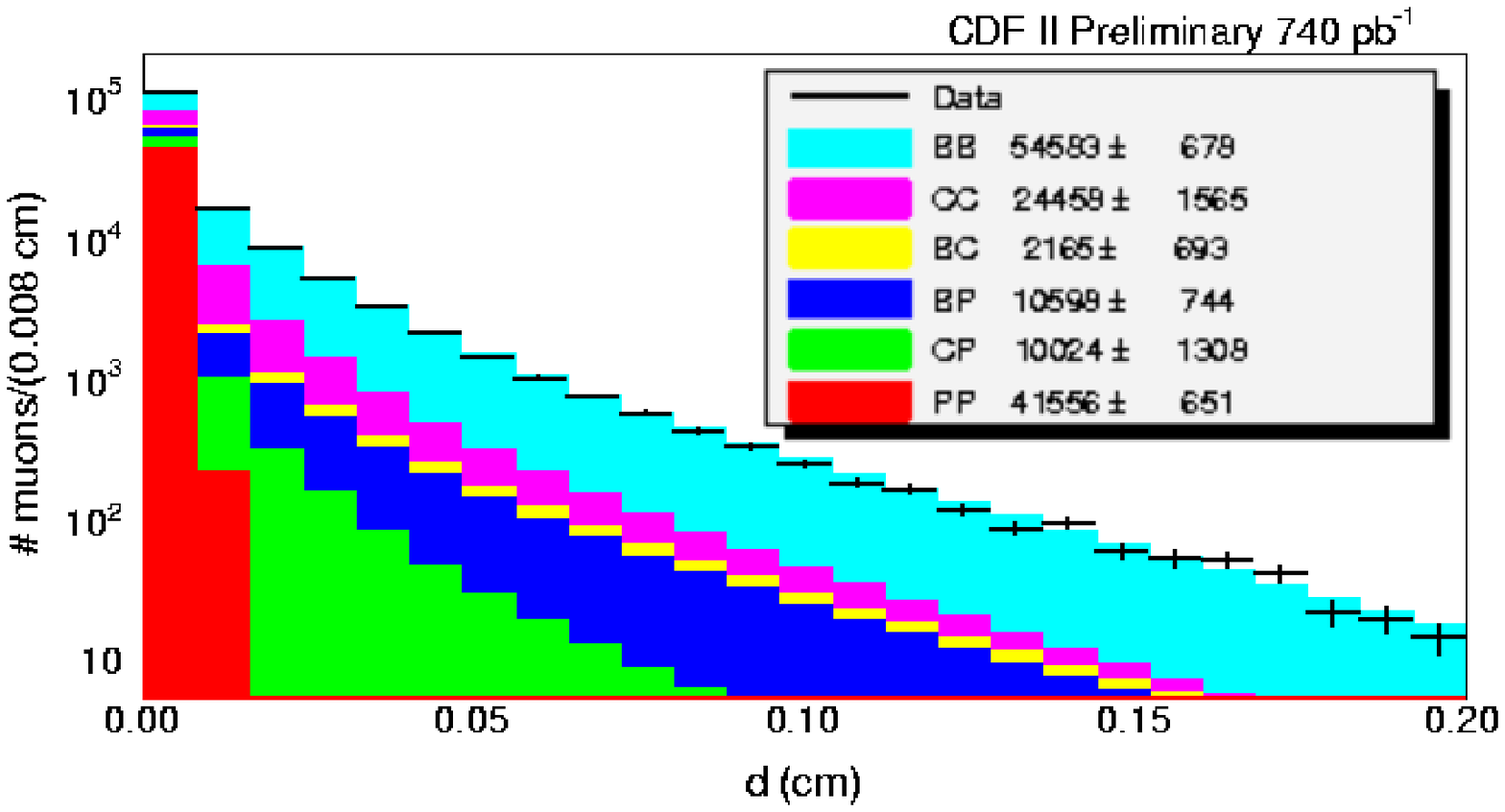,width=\linewidth}
\caption{Projection of the two-dimensional impact parameter 
	distribution of muon pairs compared to the result 
	of a f\mbox{}it for the $b\bar{b}$ and $c\bar{c}$
	components.  The different letter combinations in 
	the legend indicate the dimuon source, as explained 
	in the text: ``B'' for a muon from a $b$ quark, 
	``C'' for a muon from a $c$ quark, and ``P'' for a 
	prompt muon.  The legend gives the number of events
	found by the f\mbox{}it for each dimuon source,
	along with the statistical uncertainty.}
\label{fig:bbbar}
\end{minipage}
\hspace{0.5cm}
\begin{minipage}[tbh]{0.5\linewidth}
\centering
\psfig{figure=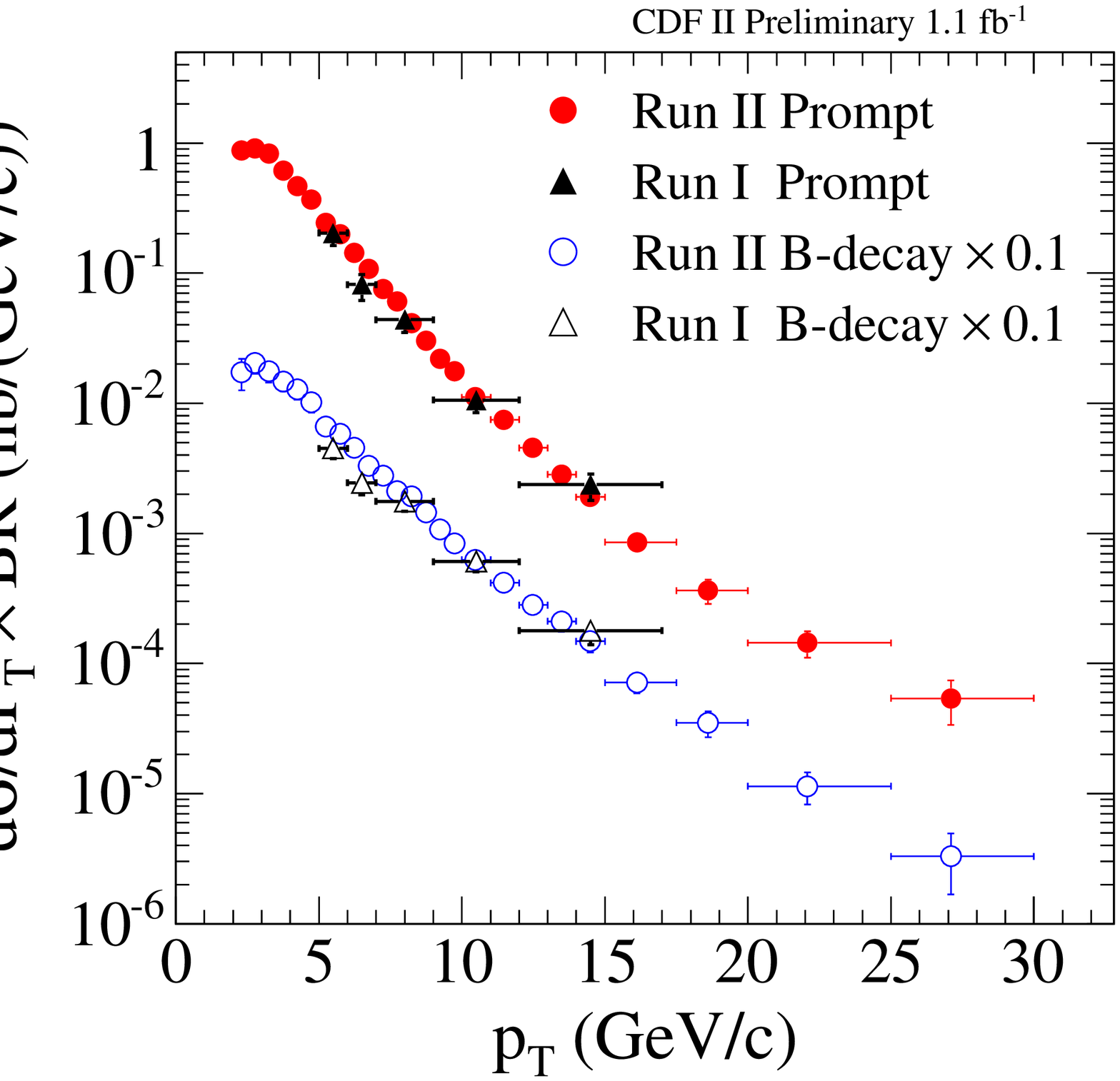,height=2.5in}
\caption{The differential cross section for prompt $\psi(2S)$
	(solid circles) and $\psi(2S)$ from $B$ meson decays (hollow circles).
	The cross section for $\psi(2S)$ from $B$ meson 
	decays is scaled down by a factor of 10 so that the two
	curves may be seen separately.  Triangular markers indicate
	the results of a CDF Run I measurement, which has not been
	scaled to account for the difference in center of mass
	energy.~\protect\cite{run1psi2s}}
\label{fig:psi2s}
\end{minipage}
\end{figure}

The f\mbox{}inal result is shown in Figure \ref{fig:psi2s},
where the promptly produced $\psi(2S)$ are separated 
from those produced in $B$ decays.  The CDF Run I
measurement is also shown in Figure \ref{fig:psi2s}.
The integrated cross section
has increased by $18\pm19$\% from the Run I measurement, 
compared to a theoretical prediction of $14\pm8$\% for
the change in center of mass energy from $1.80$ TeV to 
$1.96$ TeV.~\cite{psi2s}  The integrated inclusive
differential cross section is measured to be

\begin{tabular}{cr}
\multicolumn{2}{c}{$\sigma(p\bar{p}\to\psi(2S)X,|y(\psi(2S))|<0.6,p_{\rm T}>2{\rm GeV}/c)_{\sqrt{s}=1.96{\rm TeV}}\cdot Br(\psi(2S)\to\mu^+\mu^-)$}\\
& $= 3.17 \pm 0.04 ({\rm stat.}) \pm 0.28 ({\rm syst.})~{\rm nb}.$ \\
\end{tabular}

\section{Polarization of $\Upsilon(1S)$ and $\Upsilon(2S)$\label{sec:upsilon}}

Recent measurements of the $J/\psi$ and $\psi(2S)$
polarization at CDF show signif\mbox{}icant longitudinal
polarization with increasing $p_{\rm T}$,~\cite{psiPol}
while NRQCD predicts transverse polarization at 
suff\mbox{}iciently high $p_{\rm T}$ for $S$-wave quarkonia
produced in $p\bar{p}$ collisions.  Now D\O~presents new
measurements of the $\Upsilon(1S)$ and $\Upsilon(2S)$ 
polarization, another important test of the theoretical
approaches to QCD.

The polarization is measured in the parameter $\alpha$,
def\mbox{}ined as:
\be
\alpha=\frac{\sigma_{\rm T}-2\sigma_{\rm L}}{\sigma_{\rm T}+2\sigma_{\rm L}}
\label{eq:alpha}
\ee
where $\sigma_{\rm T}$ and $\sigma_{\rm L}$ are the 
transverse and longitudinally polarized components
of the cross section respectively.  If the transverse
and longitudinal states are equally populated, one
measures $\alpha=0$; for longitudinal polarization, 
$\alpha<0$, while for transverse polarizaton, $\alpha>0$.  
In quarkonia decay to a lepton and anti-lepton, 
$\alpha$ may be obtained from the angular distribution
\be
\frac{dN}{d(\cos\theta^*)} \propto 1 + \alpha\cos^2\theta^*
\label{eq:pol}
\ee
where $\theta^*$ is the angle between the $\Upsilon$
in the lab frame and the positive lepton in the 
$\Upsilon$ rest frame.

\begin{figure}[tbh]
\begin{center}
\psfig{figure=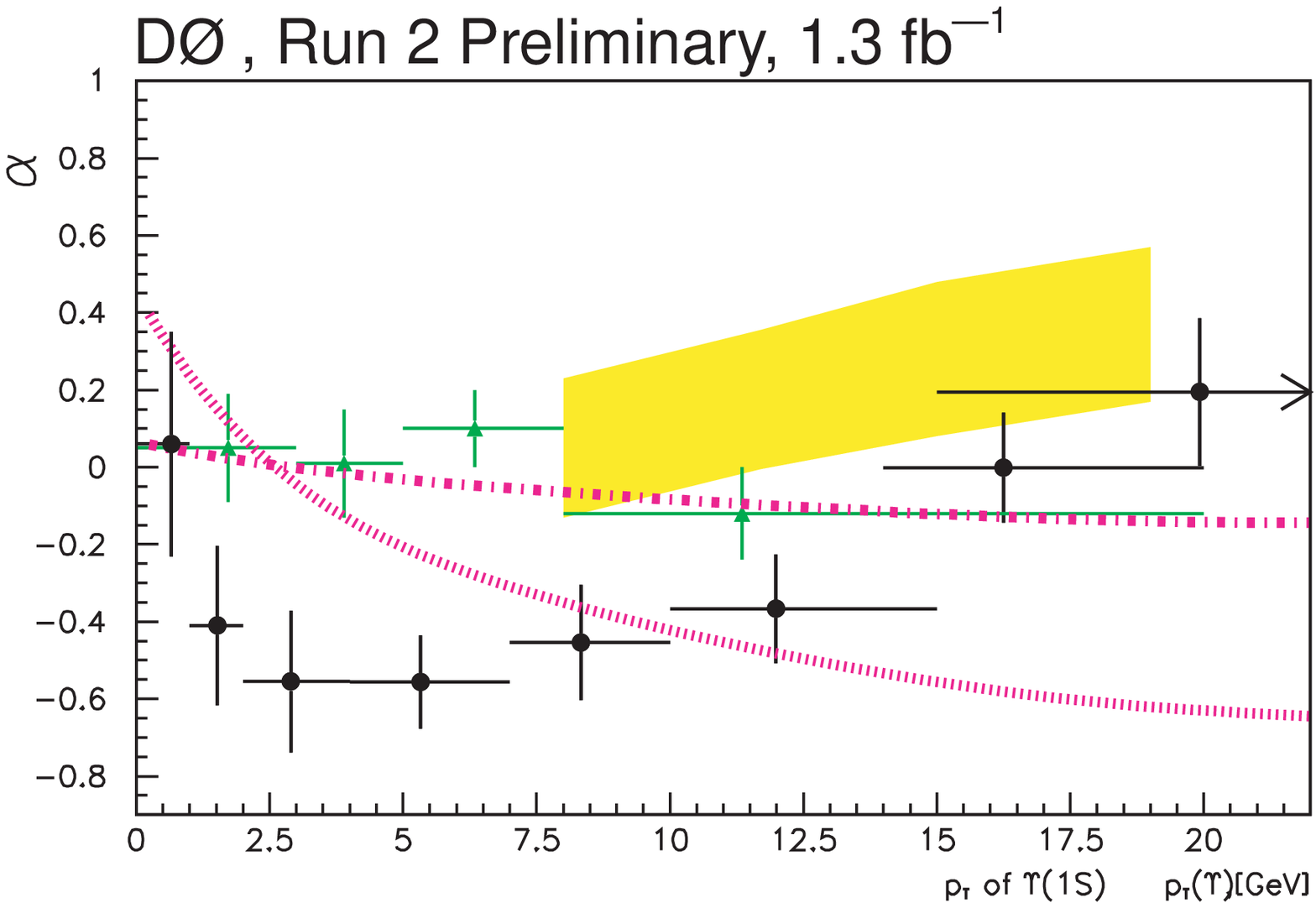,height=2.1in}
\psfig{figure=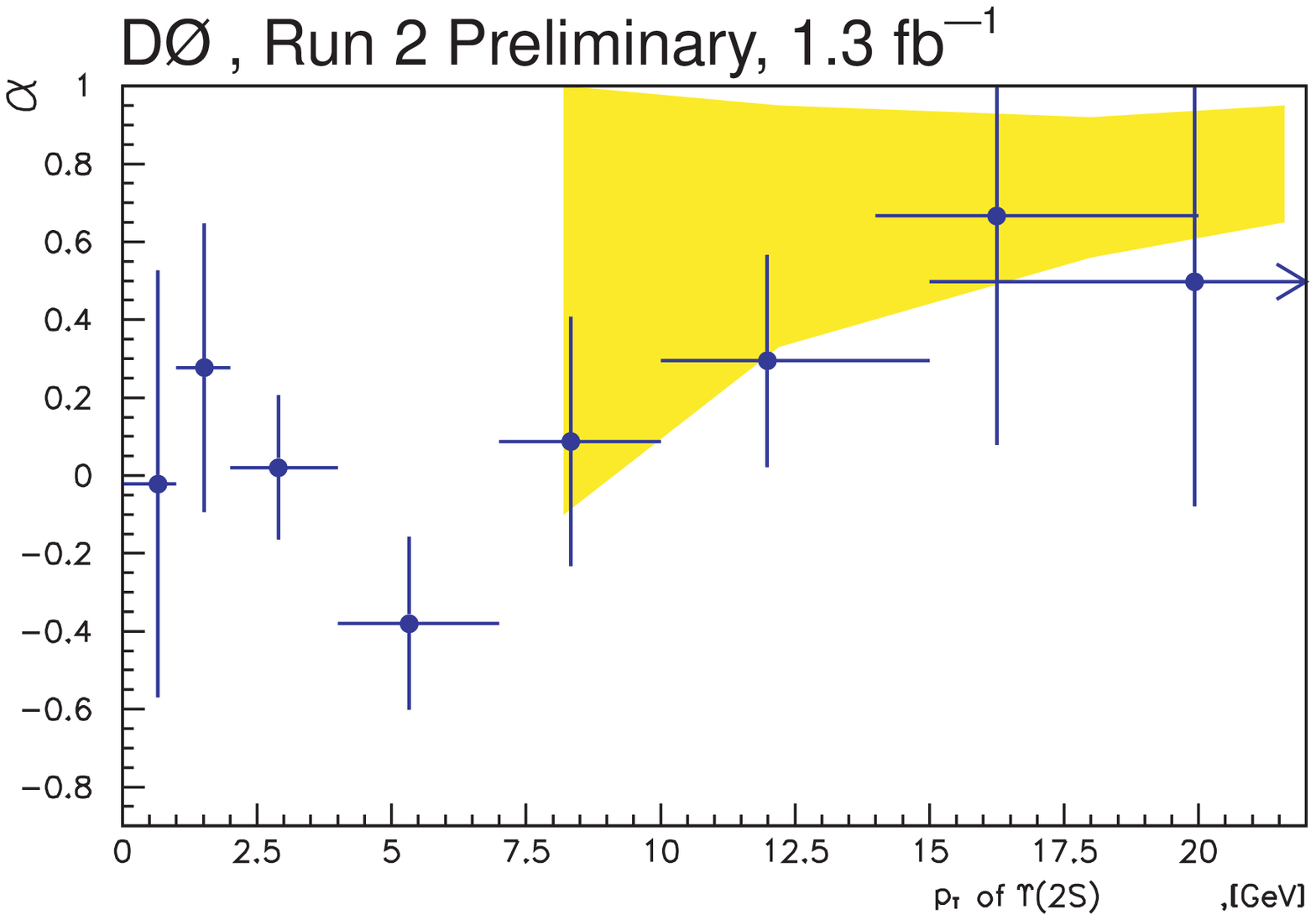,height=2.1in}
\caption{On the left is shown the dependence of the
	polarization parameter $\alpha$ on $p_{\rm T}^{\Upsilon}$
	for the $\Upsilon(1S)$.  The black points are the
	D\O~data, while the green triangles are the
	CDF Run I result.~\protect\cite{cdfupsilon}  The yellow
	band corresponds to an NRQCD prediction,~\protect\cite{ref:nrqcd}
	while the two dashed curves are two limiting cases
	from $k_{\rm T}$ factorization models.~\protect\cite{ref:kt}  
	The lower line corresponds to
	the quark-spin conservation hypothesis, while the
	upper line corresponds to the full quark-spin 
	depolarization hypothesis.  On the right is 
	shown the dependence of $\alpha$ on $p_{\rm T}^{\Upsilon}$
	for the $\Upsilon(2S)$.  The blue points are
	the D\O~data and the yellow band is an NRQCD prediction.}
\label{fig:upsilon}
\end{center}
\end{figure}

Using $1.3$ fb$^{-1}$ of D\O~data, we f\mbox{}ind 
420 000 $\Upsilon(nS) \to \mu^+\mu^-$ candidates.~\cite{upsilon} 
The $\Upsilon$ data are divided into several bins in
$\Upsilon$ $p_{\rm T}$ $(p_{\rm T}^{\Upsilon})$
and $|\cos\theta^*|$, and the number of $\Upsilon(1S)$ 
and $\Upsilon(2S)$ in each bin are extracted from 
a f\mbox{}it to the $\Upsilon$ mass distribution.
The mass signal consists of three peaks, the
$\Upsilon(1S)$, $\Upsilon(2S)$, and $\Upsilon(3S)$,
where the mass differences between the peaks are
f\mbox{}ixed to the measured values.~\cite{pdg} Unfortunately, 
the number of $\Upsilon(3S)$ was insuff\mbox{}icient to
extract the angular distributions.  The 
angular distribution in each bin is compared
to $\Upsilon(1S)$ and $\Upsilon(2S)$ Monte Carlo
samples which were generated with the parameter
$\alpha$ set to zero.  The measured value of $\alpha$ 
for the data is then determined by reweighting the
angular distributions in the Monte Carlo.  The dependence 
of $\alpha$ on $p_{\rm T}^{\Upsilon}$ is plotted in 
Figure \ref{fig:upsilon} for both $\Upsilon(1S)$ and 
$\Upsilon(2S)$, along with various theoretical
predictions.  While statistics for the $\Upsilon(2S)$ 
are insuff\mbox{}icient to draw a conclusion, in the $\Upsilon(1S)$ 
there is signif\mbox{}icant $p_{\rm T}$ dependent longitudinal
polarization which is only marginally consistent with any of
the theoretical predictions.

\section{Conclusions\label{sec:summary}}

Recent measurements of the $b\bar{b}$ correlation and
the $\psi(2S)$ cross section at the Tevatron are in 
agreement with NLO and NRQCD predictions.  However, 
measurements of quarkonia polarization in the same
perturbative $p_{\rm T}$ regime show discrepancies 
from theoretical predictions. Theoretical models
are now challenged to match the polarization
measurements while continuing to describe the cross
section data.

\section*{Acknowledgments}
We thank the Moriond QCD organizers for the invitation to
attend this conference, and are grateful for the receipt of
a travel grant from the European Union Marie Curie Program.
We also thank the Fermilab staff and the technical staffs of the 
participating institutions for their vital contributions to
this work, as well as the funding agencies supporting this work.


\begin{thebibliography}{99}
\bibitem{det_coord} The detectors use a cylindrical coordinate
        system with the $z$-axis along the proton beam direction.
        Rapidity is denoted by $y$, while the pseudorapidity
        $\eta$ is def\mbox{}ined as \mbox{$\tanh^{-1}(\cos \theta)$}.
        Transverse momentum, $p_{\rm T}$, is the component
        of the particle momentum in the $(x, y)$ plane.
\end{thebibliography}
\end{document}